\documentclass[twocolumn,aps,superscriptaddress,nolongbibliography,prl,10pt]{revtex4-2}
\usepackage{graphicx}
\usepackage{amsmath}
\usepackage{amssymb}
\usepackage{bm}
\usepackage{hyperref}
\usepackage{ulem}
\usepackage[dvipsnames]{xcolor}

\begin{document}

\title{Tunable state-dependent interactions in collisionally stable mixtures of polar molecules}
\author{Hubert J. J\'{o}\'{z}wiak}
\affiliation{Institute for Molecules and Materials, Radboud University, Nijmegen, The Netherlands}
\author{Hanwei Yang}
\affiliation{Institute for Molecules and Materials, Radboud University, Nijmegen, The Netherlands}
\author{Eugen Dizer}
\affiliation{Institut für Theoretische Physik, Universit{\"a}t Heidelberg, 69120 Heidelberg, Germany}
\author{Arthur Christianen}
\affiliation{Institute for Quantum Electronics, ETH Z{\"u}rich, 8093 Zürich, Switzerland}
\author{Tijs Karman}
\affiliation{Institute for Molecules and Materials, Radboud University, Nijmegen, The Netherlands}

\begin{abstract}
    We propose encoding a pseudo-spin-$1/2$ system in the ground ($v=0$) and first excited ($v=1$) vibrational states of polar molecules. 
    Double microwave shielding simultaneously shields molecules in both states, suppressing two-body losses by orders of magnitude while strictly avoiding three-body recombination. 
    The microwave dressing is state-dependent and results in highly tunable, long-range dipolar Ising exchange ($J_z$), density-density ($V$), and spin-density ($W$) interactions.
    These interaction length scales readily exceed the typical interparticle spacing, pushing the molecules deep into the strongly interacting regime.
    In bulk gases, this enables the exploration of itinerant quantum magnetism and quantum droplets with novel anisotropic spin textures; in optical lattices, it naturally realizes extended Hubbard and $t$-$J_z$ models, opening new directions in quantum simulation.
\end{abstract}

\maketitle

A central goal in quantum simulation is the realization of fundamental spin models~\cite{bloch2005ultracold,bloch2012quantum,Bloch2008,gross2017quantum}, such as the Hubbard~\cite{gersch,hubbard,korepin1994exactly,izyumov1995hubbard,essler2005one} and $t$-$J$ models~\cite{izyumov1997strongly,manmana,bulaevski1968new,trugman,sachdev,wrzosek1,wrzosek2025anomalouseigenstatesdopedhole}, which are believed to capture the essential physics underlying high-temperature superconductivity and magnetic ordering.
Ultracold atoms naturally realize the Hubbard model~\cite{bloch2005ultracold,bloch2012quantum,Bloch2008,gross2017quantum},
with tunneling $t$ controlled by the depth of an optical lattice,
and on-site contact interactions $U$ tunable by magnetic Feshbach resonances~\cite{Chin2010}.
The $t$-$J$ model arises as the strongly interacting limit $U\gg t$ of the Hubbard model~\cite{Chao1978},
where double occupancies occur only virtually and mediate effective off-site spin-spin interactions with strength $J\propto t^2/U$.
Consequently, these systems are limited to the $J<t$ part of the parameter regime~\cite{izyumov1997strongly}.
Furthermore, because the strongly interacting limit is achieved in deep lattices, both $t$ and $J$ are small, and the realization of low temperature compared to tunneling becomes a key bottleneck~\cite{xu2025neutral}.

Ultracold polar molecules give access to strong long-range dipolar interactions~\cite{wall2010hyperfine,trefzger2011ultracold},
offering a promising pathway to overcome these limitations by inducing strong off-site interactions directly, rather than by super-exchange~\cite{Gorshkov_2011, Gorshkov_2011_PRL}.
Quantum simulation with polar molecules is pursued experimentally~\cite{yan2013observation,rosenberg2022observation,christakis2023probing,ruttley2025long,mortlock2025multi,lu2026probing,raghuram2026probing,cornish2024quantum},
but achieving high phase-space densities and hence high lattice filling fractions has been challenging.
Remarkable progress has led to the realization of both Fermi degenerate gases~\cite{demarco:19,valtolina:20,matsuda:20,schindewolf2022evaporation} and Bose-Einstein condensates~\cite{Bigagli_2024,Shi_2026} of polar molecules.
Collisional shielding~\cite{Karman2018,Karman2025,anderegg2021observation} plays a central role by
enabling efficient evaporative cooling~\cite{Bigagli_2024}, and collisional stability~\cite{Yuan2025} with simultaneous control of molecular interactions~\cite{Karman2025,Zhang_2026}.

Gorshkov and co-authors have shown that molecules dressed in combined microwave and dc electric fields realize a fully tunable $t$-$J$ Hamiltonian extended by long-range density-density ($V$) and density-spin ($W$) interactions, referred to as the $t$-$J$-$V$-$W$ model~\cite{Gorshkov_2011, Gorshkov_2011_PRL},
and realized recently~\cite{li2023tunable,carroll2025observation}.
However, this pioneering work was carried out before it was appreciated that even non-reactive polar molecules undergo loss by ``sticky collisions''~\cite{mayle2012statistical,bause2023ultracold} and require active collisional shielding. 
Mukherjee and co-authors have shown that microwave shielding of ultracold polar molecules is compatible with pseudo-spin encoding in the hyperfine degree of freedom~\cite{mukherjee2025n}.
This results in hyperfine-independent interactions, from which SU($N$)-symmetric exchange interactions can emerge, with rich phenomenology~\cite{gorshkov2010two,taie20126,taie2022observation,zhang,hofrichter,song,surace2024scalable,werner,tusi2022flavour,cazalilla2014ultracold,ibarra2025many}.
Yet, because these effective interactions arise indirectly through quantum statistics, they arise on prohibitively small energy scales.
Thus, a unified platform that simultaneously provides strong, directly tunable state-dependent interactions has yet to be realized.

Here, we propose to encode pseudo-spin in the ground ($v=0$) and the first excited ($v=1$) vibrational states of microwave-shielded polar molecules.
We show that this leads to a collisionally stable multi-component gas with long-range anisotropic dipolar interactions that are tunably spin-dependent.
Using bosonic NaCs as a highly relevant example, we show this setup provides broad tunability and access to the strongly interacting regime, while essentially eliminating collisional loss. Crucially, realizing this platform requires minimal experimental overhead~\footnote{In practice, populating the $v=1$ state can be achieved by utilizing the stimulated Raman adiabatic passage (STIRAP) lasers already present for ground-state preparation, requiring only a single additional laser to drive the transition into the vibrationally excited state. No additional microwave fields are required compared to shielding of a single vibrational state.}, relying entirely on state-preparation and microwave shielding techniques that are already standard in modern ultracold molecule setups~\cite{Bigagli_2023}.

\begin{figure}
    \centering
    \includegraphics[width=\linewidth]{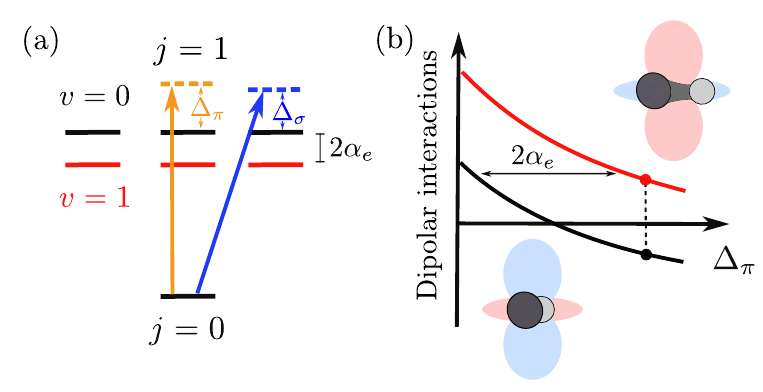}
    \caption{{\bf State-dependent microwave shielding and tunable dipolar interactions}. (a) Energy-level diagram for the $v=0$ (black) and $v=1$ (red) vibrational states. Microwave fields ($\pi, \sigma$) couple the $j=0$ and $j=1$ rotational states. Rotation-vibration coupling effectively increases the microwave detunings ($\Delta_{\sigma,\pi}$) for the $v=1$ state by $2\alpha_e$. (b) The resulting effective dipolar interactions as a function of $\Delta_{\pi}$. The $2\alpha_e$ difference shifts the $v=1$ (red) curve relative to the $v=0$ curve (black), enabling state-dependent dipolar interactions between the ground- and excited-state molecules. The cartoons illustrate the anisotropy of these interactions and the physical effect of the sign reversal between dipolar and anti-dipolar interactions.}
    \label{fig:1}
\end{figure}

We consider a mixture of polar molecules in the vibrational ground ($v=0$) and first excited ($v=1$) states under double microwave shielding~\cite{Karman2025}, see Fig. \ref{fig:1}.
The gas is globally dressed by $\sigma^+$- and $\pi$-polarized microwave fields near-resonant with the rotational $j=0\to 1$ transition,
characterized by their Rabi frequencies ($\Omega_{\sigma}, \Omega_{\pi}$) and detunings ($\Delta_{\sigma}, \Delta_{\pi}$).
The rotational transition frequency is $v$-dependent,
leading to an effective detuning $\Delta^{(v)}_{\sigma,\pi} = \Delta^{(0)}_{\sigma,\pi} + 2\alpha_e v$ where the rotation-vibration coupling constant $\alpha_e$ is typically on the order of MHz~\cite{rvdw}.
Since microwave Rabi frequencies and detunings typical for shielding are in the order of tens of MHz~\cite{anderegg2021observation,schindewolf2022evaporation,Bigagli_2023,Bigagli_2024,lin2023microwave,Shi_2026},
the same microwave fields can simultaneously dress molecules in different vibrational states.
However, the dressing is not identical in different vibrational states,
meaning the resulting dipolar interaction strength scales as $(1+\left[\Delta^{(v)}/\Omega\right]^2)^{-1}$, see Fig.~\ref{fig:1}.
The central result of this work is that this dressing can simultaneously induce appreciably state-dependent interactions and suppress collisional loss in all relevant vibrational states.

The strength of the dipolar interactions,
\begin{align}
V_\mathrm{dd}^{v,v'} = a_{dd}^{v,v'} \frac{\hbar^{2}}{2M} \frac{1-3\cos^2\theta_{12}}{R_{12}^3},
\end{align}
is parameterized by the dipolar lengths $a_{dd}^{v,v'}$ which in general depend on the vibrational state of both molecules $v,v'$.
This pseudo-spin-dependence can equivalently be written as
\begin{align}
V_\mathrm{dd} = \frac{\hbar^{2}}{2M} \frac{1-3\cos^2\theta_{12}}{R_{12}^3} \left[J_z \hat{S}^z_{1} \hat{S}^z_2 + V + W \left(\hat{S}^{z}_1 +\hat{S}^{z}_2\right)  \right],
\end{align}
with the Ising $J_z = a_{dd}^{00} + a_{dd}^{11} - 2a_{dd}^{01}$, 
density-density $V = (a_{dd}^{00} + a_{dd}^{11} + 2a_{dd}^{01})/4$,
and the density-spin coupling $W = (a_{dd}^{00} - a_{dd}^{11})/2$.
We emphasize that $J_z$, $V$, and $W$ are dipolar length scales.
The interactions are long-ranged and anisotropic,
which can be different from the usage in typical extended $t$-$J$ models defined on a lattice.
Depending on the polarization direction of the effective dipoles,
the interactions can be made isotropic in a two-dimensional lattice, as typically considered,
but anisotropic off-site interactions are also possible.
By controlling the microwave polarization, the type of anisotropy can also be tuned~\cite{Zhang_2026,chen2023field}.
We note that the scattering lengths that dictate contact interactions in dilute Bose gases are also $v$-dependent and tunable, as we show in Fig.~\ref{fig:tunability_scattering_length}.
This gives further opportunities for interaction control, but we focus on the control of dipolar interactions.

We note that the molecules experience no transverse ``flip-flop'' interactions [$J_\perp (\hat{S}^x_{1} \hat{S}^x_2 + \hat{S}^y_{1} \hat{S}^y_2)$],
which in principle result from first-order dipole-dipole interactions between molecules in different vibrational states.
This interaction is proportional to the square of the off-diagonal vibrational transition dipole moment, $\langle v=0|\hat{d}|v=1\rangle$.
However, for typical ultracold molecules, this transition dipole is much smaller than the permanent dipole moment ($d_v$),
and this interaction is negligible~\cite{rvdw}.

\begin{figure}
    \centering
    \includegraphics[width=\linewidth]{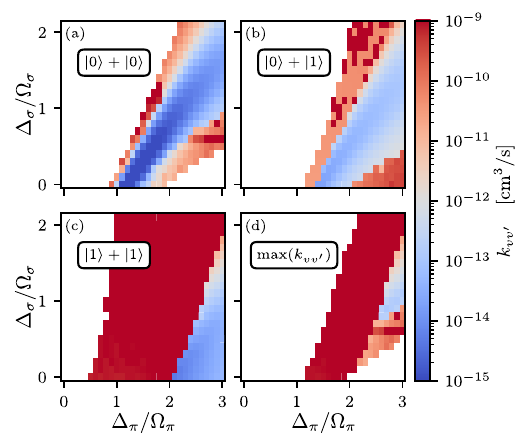}
    \caption{{\bf Two-body loss rates} for bosonic NaCs, derived from the complex $s$-wave scattering length (see Appendix~\protect\hyperlink{app:a}{A}). The rates are computed at the optimal overall field strength below $\Omega^{\mathrm{tot}} = 50\times 2\pi\,\mathrm{MHz}$ for a $\pi$-dominated field configuration ($\Omega_{\sigma}/\Omega_{\pi} = 1/4$). Panels (a--c) show the optimized loss rates for collisions between identical ground ($|0\rangle+|0\rangle$), distinguishable ($|0\rangle+|1\rangle$), and identical excited ($|1\rangle+|1\rangle$) molecules, respectively. Panel (d) displays the maximum of these three loss rates, $\mathrm{max}(k_{vv'})$. The white regions are excluded due to the occurrence of field-linked bound states.}
    \label{fig2}
\end{figure}

The molecular mixture must first be collisionally stable with respect to both two-body and three-body loss.
For all possible pairs of vibrational states ($|v\rangle$ + $|v'\rangle$) -- identical ground ($|0\rangle+|0\rangle$), identical excited ($|1\rangle+|1\rangle$), and distinguishable ($|0\rangle+|1\rangle$) pairs -- we identify microwave parameters that minimize two-body collisional loss while preventing formation of field-linked bound states,
which would otherwise result in three-body recombination~\cite{stevenson2024three,Yuan2025}.
To reduce the parameter space, we follow Ref.~\cite{map} and consider that at any fixed relative field strength, $\Omega_{\sigma}/\Omega_{\pi}$, and fixed detuning ratios, $\Delta_{\sigma}/\Omega_{\sigma}$ and $\Delta_{\pi}/\Omega_{\pi}$,
a bound state emerges as one increases $\Omega^{\mathrm{tot}} = \sqrt{\Omega_{\sigma}^{2}+\Omega_{\pi}^{2}}$.
Hence, the occurrence of the first bound state for any pair of vibrational states, or a practical upper limit of $50\times 2\pi$~MHz, will set a maximum Rabi frequency,
and below it there will be a unique Rabi frequency that also minimizes collisional loss. We refer to the corresponding minimum rate as the {optimized} loss rate.

\begin{figure}
    \centering
    \includegraphics[width=\linewidth]{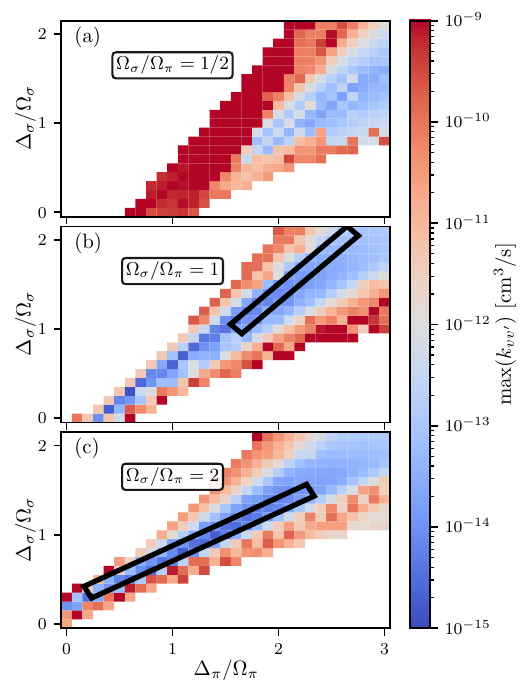}
    \caption{{\bf The maximum two-body loss rate for various Rabi frequency ratios}. The overall field strength is capped at $\Omega^{\mathrm{tot}} = 50\times 2\pi\,\mathrm{MHz}$. Marked in bold are the collisionally stable regions that we investigate in more detail in Fig.~\ref{fig4}.}
    \label{fig3}
\end{figure}

Using coupled-channel quantum scattering calculations (see Appendix~\hyperlink{app:a}{A}), we extract these optimized loss rates---$k_{00}$, $k_{11}$, and $k_{01}$---from the imaginary part of the scattering length. Figure~\ref{fig2}(a-c) shows the resulting values for all three collisional pairs as a function of detuning ratios, $\Delta_{\sigma}/\Omega_{\sigma}$ and $\Delta_{\pi}/\Omega_{\pi}$, for a $\pi$-dominated field configuration ($\Omega_{\sigma}/\Omega_{\pi} = 1/4$). The regions of deep loss suppression shift towards larger $\pi$-detunings for collisions involving vibrationally excited molecules.
Because the experimental mixture is subjected to a single, global microwave configuration, the overall stability of the gas is dictated by the highest of these three loss rates, $\max(k_{vv'})$. Evaluating this maximum loss rate (Fig.~\ref{fig2}(d)) reveals a viable overlap window at $\Delta_{\pi}/\Omega_{\pi} > 2.5$, where losses across all three collisional pairs are simultaneously suppressed.

We systematically map the maximum loss rate, $\mathrm{max}(k_{vv'})$, across different relative field strengths ($\Omega_{\sigma} / \Omega_{\pi}$), as shown in Fig. \ref{fig3}. We find that balanced ($\Omega_{\sigma} / \Omega_{\pi} = 1$) and $\sigma$-dominated ($\Omega_{\sigma} / \Omega_{\pi} = 2$) configurations feature significantly broader stability windows, allowing deeper simultaneous suppression of loss. Crucially, we verify that within these windows, vibrationally inelastic processes, such as relaxation in $|0\rangle+|1\rangle$ collisions, provide a negligible contribution to the total loss rate (see Appendix~\hyperlink{app:c}{C}).

\begin{figure}
    \centering
    \includegraphics[width=\linewidth]{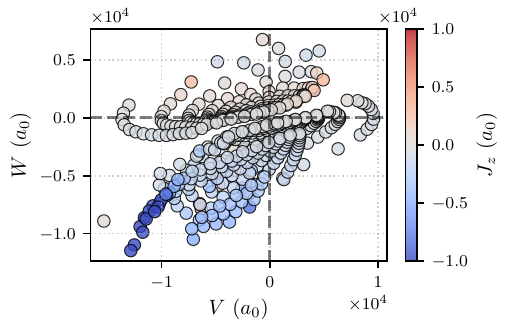}
    \caption{{\bf Accessible range of the density-density ($V$), density-spin ($W$) and Ising interactions} ($J_{z}$, see the color scale) for arbitrary microwave configuration where the maximum loss rate, $\mathrm{max}(k_{vv'})$, remains below $10^{-12}$ cm$^{3}$/s.
    }
    \label{fig:WVJmap}
\end{figure}

Having established broad parameter regimes of simultaneous collisional stability, we now demonstrate that scanning through these provides independent control over the interaction parameters.
To gauge the range of tunability of the interaction parameters, Fig.~\ref{fig:WVJmap} shows the Ising ($J_{z}$), density-density ($V$), and density-spin ($W$) realizable with loss rate coefficients below $10^{-12}$~cm$^{3}$/s.
Interactions of all types can be strong,
with dipolar lengths on the 10\,000~$a_0$ scale,
and $J_z,\,V$ can be realized independently with either sign.

\begin{figure}
    \centering
    \includegraphics[width=\linewidth]{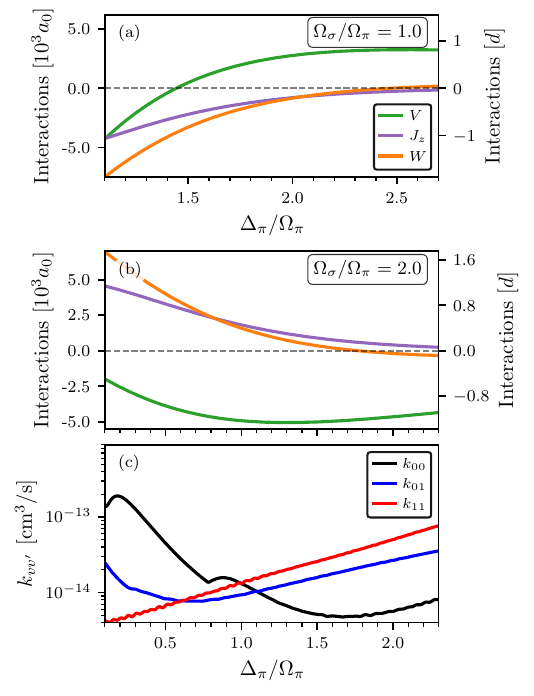}
    \caption{{\bf Tunability of the model parameters along collisionally stable paths.} {(a, b)} Ising ($J_z$), density-density ($V$), and density-spin ($W$) interactions along the two trajectories shown in Fig.~\ref{fig3}(b) and Fig.~\ref{fig3}(c), respectively.
    The right vertical axes indicate the interaction strengths scaled by the interparticle distance $d \approx  4000\ a_0$, corresponding to a bulk gas density of $10^{14}\ \text{cm}^{-3}$. {(c)} The corresponding two-body loss rates along the trajectory shown in panel (b).
    The loss rates for the trajectory in panel (a) exhibit similar suppression below $10^{-13}\,\mathrm{cm}^{3}/\mathrm{s}$ (not shown).
    The overall field strength is fixed at $\Omega^{\mathrm{tot}} = 20\times 2\pi\,\mathrm{MHz}$.}
    \label{fig4}
\end{figure}

We highlight two compelling tuning paths starting from configurations where the spin-dependence vanishes ($W \approx 0, J_{z} \approx 0$), but the spin-independent density-density interaction ($V$) is strong.
Where the spin-density interaction $W$ vanishes, the system exhibits SU(2) symmetry.
From these points, we can tune the system into spin-dependent regimes with substantial Ising spin-spin ($J_z$) and a comparable or dominant spin-density ($W$) dipolar interaction.
The first path occurs for balanced fields ($\Omega_{\sigma}/\Omega_{\pi} = 1$) and is shown in Fig.~\ref{fig3}(b).
The resulting interactions are shown Fig.~\ref{fig4}(a).
At $\{\Delta_{\sigma}/\Omega_{\sigma}, \Delta_{\pi}/\Omega_{\pi}\} = \{2.1, 2.7\}$,
the spin dependence of the interaction is negligible.
By sweeping the $\pi$-detuning down, strong anti-ferromagnetic Ising couplings ($J_{z} < 0$) and prominent density-spin interactions emerge.
The second path occurs for {($\Omega_{\sigma}/\Omega_{\pi} = 2$)} as shown in Fig.~\ref{fig3}(c).
In this case, the interactions can be tuned from spin-independent to strongly ferromagnetic Ising coupling ($J_z > 0$) and a strong spin-density interaction, see Fig. \ref{fig4}(b).

Crucially, this wide parameter tunability does not come at the cost of the stability of the gas. As shown in Fig.~\ref{fig4}(c), the two-body loss rates for all three distinct types of collisions ($k_{00}$, $k_{11}$, and $k_{01}$) remain continuously suppressed below $10^{-13}$ cm$^{3}$/s.


To gauge the strength of the achievable interactions, parameterized as dipolar length scales $J_z$, $V$, $W$, we compare them to the interparticle spacing in a bulk gas, $d \equiv n^{-1/3}$.
The ratio $a_{dd}/d$ is interpreted as the ratio of the mean dipolar interaction energy to the mean kinetic energy in a non-interacting gas.
In a bulk gas, inducing these dipolar interactions naturally drives the formation of quantum droplets~\cite{Zhang_2026, Shi_2026, jung2026staticfieldshieldingbosonicmolecules}, where densities can reach $n \approx 10^{14}$~cm$^{-3}$, corresponding to the interparticle distance of $d\approx 4 \times 10^{3} \, a_{0}$.
As can be seen in Fig.~\ref{fig4}(a, b), both paths allow tuning interaction length scales exceeding this distance. 
Combined with the loss rates suppressed below $10^{-13}$~cm$^{3}$/s, this yields two-body lifetime exceeding $100\,\mathrm{ms}$, which is sufficient to observe many-body dynamics.
Away from these two trajectories, the interaction length scales can even exceed 10,000~$a_0 \approx 500$~nm, reaching typical lattice constants and maintaining the strongly interacting regime even at lower densities, see Figs.~\ref{fig:WVJmap} and \ref{fig:NaCs_lowerdensities}. For ultrapolar molecules consisting of alkali-metal and coinage-metal atoms~\cite{Smialkowski2021}, the interaction length scales can be several times larger~\cite{Dutta_2025,map}, as illustrated in Fig.~\ref{fig:KAg}, pushing further into the strongly interacting regime. 

Although in this manuscript we have focused on bosonic NaCs, the proposed shielding framework applies equally to fermionic molecules.
Fermionic mixtures are expected to be at least as robust against collisional loss,
and the induced dipolar interactions are independent of quantum statistics.
The ability to induce strong, tunably state-dependent dipolar interactions therefore extends directly to multi-component Fermi degenerate gases of polar molecules.
Microwave shielding in single-component gases can already lead to unconventional $p$-wave superfluidity~\cite{baranov:2002,cooper:2009,levinsen:2011,deng:2023},
and the extension to tunable state-dependent interactions can be used to drive competing pairing mechanisms.

Our model can be extended by incorporating additional internal degrees of freedom. 
The vibrational pseudo-spin can be combined with $N$ nuclear hyperfine states in the ground and excited manifolds, analogous to two-orbital SU($N$) magnetism in alkaline-earth atoms~\cite{gorshkov2010two}.
Under the conservation of vibrational excitations, this setup exhibits a $\mathrm{U}(1)\times \mathrm{SU}(N)\times \mathrm{SU}(N)$ symmetry, which can be extended to the full $\mathrm{SU}(2N)$ symmetry at the specific microwave parameters where the spin-dependent interactions vanish ($J_z=W=0$). Additionally, our scheme is not limited to the $v=0$ and $v=1$ states and can be extended to higher vibrational manifolds (see Appendix~\hyperlink{app:d}{D}) to provide even more degrees of freedom.

In conclusion, we have proposed encoding pseudo-spin into the vibrational degree of freedom of ultracold polar molecules, where double microwave shielding eliminates collisional loss. This realizes long-range anisotropic dipolar interactions of the Ising exchange, density-density, and spin-density type that are all tunable by the microwave fields.
The detunings and Rabi frequencies of the two microwave fields provide flexible independent control over interactions in multiple states, unlike for atoms where all interactions are tuned by a single magnetic field~\cite{Bloch2008, Chin2010}.
The interaction length scales can exceed typical interparticle spacing in both optical lattices and bulk gases.

In bulk gases, this corresponds to the strongly interacting regime~\cite{Yuan2025,schindewolf2025few}, where already in the single-component case the dipolar interaction drives formation of quantum droplets, as observed in~\cite{Zhang_2026}, and competing phases are predicted~\cite{ciardi2025self}. Although our setup realizes a two-component pseudo-spin mixture rather than a spinor Bose-Einstein condensate~\cite{stamperkurn,silvera1986spin,guirk}, it offers a distinct mechanism for pattern formation. Specifically, the dipolar spin-density interaction found here favors an anisotropic spin texture in inhomogeneous gases and droplets---a phenomenon that in degenerate Bose gases cannot arise from exchange interactions alone~\cite{stamperkurn}.

In the lattice setting, the dipolar interactions realize long-range extended Hubbard models with tunably spin-dependent off-site interactions~\cite{koinov,Harir23032016}. 
In the limit of hard-core repulsion that prevents double occupancy, this realizes an extension of the $t$-$J$ model~\cite{izyumov1997strongly,manmana}---a paradigmatic framework for strongly correlated electrons that hosts rich physics for Ising exchange~\cite{bulaevski1968new,trugman,sachdev,wrzosek1,wrzosek2025anomalouseigenstatesdopedhole}.
In the extended model, long-range density-density interactions are predicted to enhance superfluid correlations~\cite{troyer1993spin},
and the Ising spin-spin and spin-density interactions can induce phase separation~\cite{mao2015phase} or enhance $d$-wave superfluidity~\cite{kuns}. 

\textit{Acknowledgements---} We thank Sebastian Will, Ian Stevenson, Holly Middleton-Spencer, Kaden Hazzard, and Richard Schmidt for useful discussions.
The research was funded by the European Union (Project No.~101269084, HORIZON-MSCA-2025-PF, 2STICKY). The views and opinions expressed are, however, those of the authors only and do not necessarily reflect those of the European Union or the European Research Executive Agency. Neither the European Union nor the granting authority can be held responsible for them.
H.Y.~and T.K.~are supported by NWO VIDI (grant ID 10.61686/AKJWK33335).

\newpage
\section{End Matter}

\hypertarget{app:a}{\textit{Appendix A: Quantum scattering calculations}} --- Loss rate coefficients are obtained from coupled-channel quantum scattering calculations that include a short-range absorbing boundary condition~\cite{Karman2018, Karman2019, Karman2020, anderegg2021observation, schindewolf2022evaporation, chen2023field, Bigagli_2023, Bigagli_2024, Karman2025, rvdw}, which models collisional loss due to sticky collisions~\cite{mayle2012statistical,Mayle2013,Christianen2019a,bause2023ultracold}. Details of these calculations are discussed in Ref.~\cite{Karman2025}, while the extension of the approach to include vibrational degrees of freedom is presented in Ref.~\cite{rvdw}. We extract the two-body loss rate coefficients from the imaginary part of the complex $s$-wave scattering length, $a_{s}^{v,v'} = \alpha_{s}^{v,v'} - i\beta_{s}^{v,v'}$, via
\begin{equation}
    k_{vv'} = \frac{4 \pi \hbar }{\mu} g_{vv'} \beta_{s}^{v,v'},
\end{equation}
where $\mu$ is the reduced mass of the scattering system and $g_{vv'}$ is the degeneracy factor, which equals 2 for identical molecules ($v=v'$) and $1$ for distinguishable molecules ($v\neq v'$).

\begin{figure}[!ht]
    \centering
    \includegraphics[width=\linewidth]{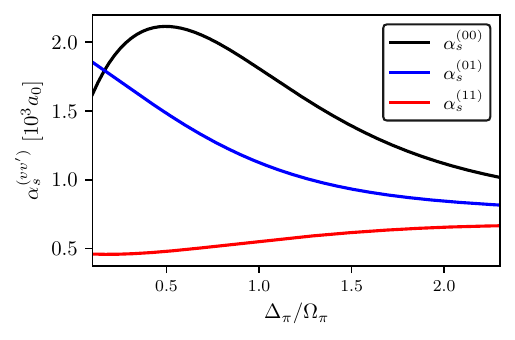}
    \caption{{\bf Tunability of the real part of the scattering lengths, $\alpha_{s}^{vv'}$}, along the trajectory shown in Fig.~\ref{fig3}(c). The overall field strength is fixed at ${\Omega^{\mathrm{tot}} = 20\times 2\pi\,\mathrm{MHz}}$.}
    \label{fig:tunability_scattering_length}
\end{figure}

\hypertarget{app:b}{\textit{Appendix B: Tunability of contact interactions}}
 --- Beyond suppressing two-body losses and inducing state-dependent dipolar interactions, the proposed scheme allows extensive control over the contact interactions. As noted in the main text, the real parts of the $s$-wave scattering lengths, $a_{s}^{v,v'} = \alpha_{s}^{v,v'} - i\beta_{s}^{v,v'}$, are state-dependent and tunable via applied microwave fields. Figure \ref{fig:tunability_scattering_length} illustrates this tunability for the $|0\rangle + |0\rangle$, $|0\rangle + |1\rangle$, and $|1\rangle + |1\rangle$ collisions along the trajectory shown in Fig.~\ref{fig3}(c) ($\Delta_{\sigma}/\Omega_{\sigma} = \frac{6}{11}\Delta_{\pi}/\Omega_{\pi} + \frac{27}{110}$,  $\Omega_\sigma/\Omega_\pi = 2$).
 Crucially, this control over the contact interactions is achieved without compromising the stability of the gas, as the corresponding two-body loss rates remain suppressed below the $10^{-13}\,\mathrm{cm}^3\mathrm{/s}$ level (see Fig.~\ref{fig4}(c)).

\hypertarget{app:c}{\textit{Appendix C: Vibrationally inelastic processes}} --- 
To generate state-dependent interactions, the microwave detunings must be comparable to the rotation-vibration coupling constant ($2\alpha_e$). In this regime, the composition of the field-dressed states, which determines shielding efficiency and the induced dipolar interactions, is sensitive to the precise microwave parameters. Because of this sensitivity, it is not \textit{a priori} clear whether the two-body loss rates can be simultaneously suppressed for all collision pairs ($|0\rangle + |0\rangle$, $|0\rangle + |1\rangle$ and $|1\rangle + |1\rangle$), while ensuring the absence of the field-linked bound states that lead to three-body recombination.

Furthermore, the presence of vibrational degrees of freedom introduces new inelastic collision channels that could potentially lead to higher loss rates. For distinguishable $|0\rangle + |1\rangle $ collisions, vibrational relaxation to $|0\rangle + |0\rangle $ releases one vibrational quantum of energy, which is on the order of THz for bialkali molecules. 
By contrast, identical $|1\rangle+|1\rangle $ collisions can undergo a near-resonant transition to $|0\rangle + |2\rangle$, which releases an energy determined by the vibrational anharmonicity (typically on the order of GHz for bialkali molecules). 

\begin{figure}
    \centering
    \includegraphics[width=\linewidth]{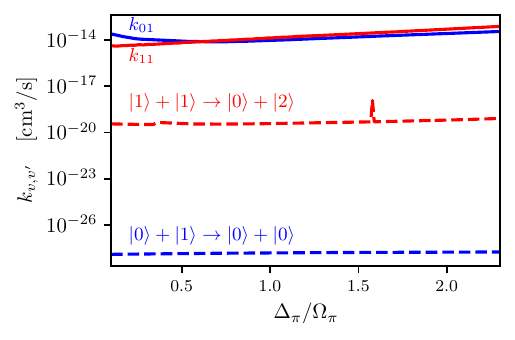}
    \caption{{\bf Contribution of the vibrationally inelastic processes}, $|0\rangle + |1\rangle \to |0\rangle + |0\rangle$ and $|1\rangle + |1\rangle \to |0\rangle + |2\rangle$, to to the total two-body loss rate for distinguishable ($k_{01}$) and identical excited ($k_{11}$) molecular collisions along the trajectory shown in Fig.~\ref{fig3}(c). The overall field strength is fixed at $\Omega^{\mathrm{tot}} = 20\times 2\pi\,\mathrm{MHz}$.}
    \label{fig:vib_relax}
\end{figure}

To quantify these processes, Fig.~\ref{fig:vib_relax} compares the total loss rate against the state-to-state rates for the $|0\rangle + |1\rangle \to |0\rangle + |0\rangle$ and $|1\rangle + |1\rangle \to |0\rangle + |2\rangle$ transitions, along the $\Omega_\sigma/\Omega_\pi = 2$ path shown in Fig.~\ref{fig3}(c). We find that the vibrationally inelastic contributions are suppressed by many orders of magnitude relative to the total loss rate, confirming that vibrational relaxation is negligible and does not deteriorate shielding efficiency.

\hypertarget{app:d}{\textit{Appendix D: Extension to higher vibrational manifolds}} --- While the main text focuses on the $v=0$ and $v=1$ states, the shielding scheme is not limited to this two-level subspace. Since vibrationally inelastic processes are negligible (Fig.~\ref{fig:vib_relax}), the vibrational quantum number enters the Hamiltonian essentially only through the effective rotational constant, $B_v = B_e - \alpha_e(v+1/2)$. Consequently, using a high vibrationally excited state (at a fixed $\alpha_e$) is equivalent to operating with a proportionally larger rotation-vibration coupling constant (at fixed $v=1$). Although we do not explicitly perform scattering calculations for $v > 1$, this indicates that the scheme can simultaneously work in higher vibrational manifolds. This is particularly promising for heavier molecules with significantly smaller rotation-vibration coupling constants. For instance, in RbAg and CsAg, the coupling constants are $\alpha_{e} =3.4$ and $2.0\,\mathrm{MHz}$, respectively, compared to $\alpha_{e}=7.09$ and $8.1\,\mathrm{MHz}$ for NaCs and KAg~\cite{rvdw}.
This suggests that the state-dependent shielding scheme could work simultaneously across $v=0-4$.

\begin{figure}[!hb]
    \centering
    \includegraphics[width=\linewidth]{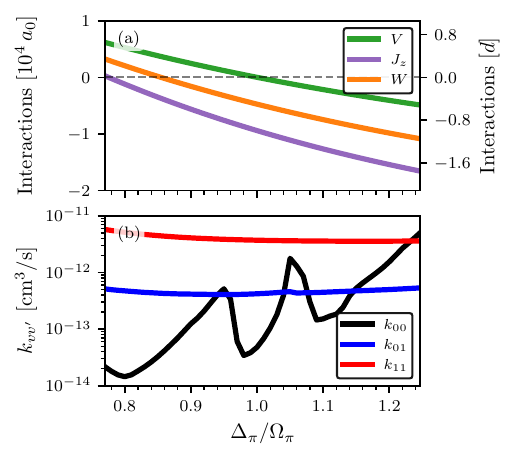}
    \caption{{\bf (a) Tunability of the interaction model parameters ($V$, $J_z$, $W$) and (b) the two-body loss rates ($k_{00}$, $k_{01}$, $k_{11}$)} along the trajectory defined by $\Omega_{\sigma}/\Omega_{\pi} = 1$, $\Delta_{\sigma}/\Omega_{\sigma} = 0.5$, $\Delta_{\pi} / \Omega_{\pi} \in [0.75-1.25]$. The right vertical axis in (a) displays the interaction strengths scaled by the interparticle distance $d \approx 10\,000\,a_0$, corresponding to a density of $n = 8 \times 10^{12}\,\mathrm{cm}^{-3}$. The overall field strength is fixed at $\Omega^{\mathrm{tot}} = 5\times 2\pi\,\mathrm{MHz}$.}
    \label{fig:NaCs_lowerdensities}
\end{figure}

\textit{Appendix E: Interaction strengths at lower gas densities} --- In the main text, we gauged the interaction length scales assuming the density of a quantum droplet ($n\approx 10^{14}\,\mathrm{cm}^{-3}$). 
To demonstrate that our shielding scheme does not require this,
we consider a reduced density of $n\approx 8\times 10^{12}\,\mathrm{cm}^{-3}$ and corresponding interparticle spacing $d \approx 10,000\,a_{0}$.
This is roughly representative of a dilute quantum gas and also coincides with the interparticle spacing in a 1064~nm optical lattice at unit filling.
Figure \ref{fig:NaCs_lowerdensities} shows an alternative tuning trajectory that allows tuning $J_{z}$ up to $\approx 1.7\times10^{4}\,a_{0}$, nearly a factor of two larger than the lattice spacing. Along this path, the two-body loss rate remains suppressed below $5\times 10^{-12}\,\mathrm{cm}^{3}/\mathrm{s}$. This yields a two-body lifetime exceeding 25 ms, which is sufficient to observe many-body dynamics.

\hypertarget{app:e}{\textit{Appendix F: Application to ultrapolar molecules}} --- 
Ultrapolar silver-bearing molecules possess significantly larger permanent dipole moments ~\cite{Smialkowski2021}, which directly boost the achievable interaction length scales, since $a_{dd} \propto d^2$. However, simultaneous loss suppression across multiple vibrational states is not guaranteed \textit{a priori}. The strong dependence of the characteristic dipolar energy $E_{d}\propto d^{-4}$ alters the relevant energy scales of these systems. First, the ratio of the rotation-vibration coupling constant to the dipolar energy, $\alpha_{e}/E_{d}$, becomes orders of magnitude larger~\cite{rvdw}, which influences the collision dynamics involving excited vibrational states. Second, this shifted energy scale means that the two-body bound states emerge at lower Rabi frequencies \cite{Dutta_2025,map}, which could limit the optimal operating regimes identified in Figs.~\ref{fig2} and~\ref{fig3}. 

\begin{figure}
    \centering
    \includegraphics[width=\linewidth]{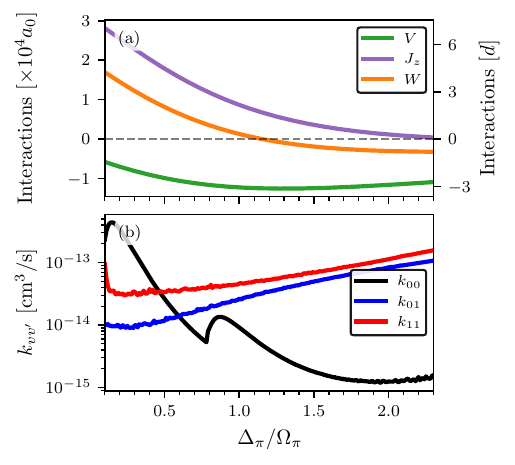}
    \caption{{\bf (a) Tunability of the interaction model parameters ($V$, $J_z$, $W$) and (b) the two-body loss rates ($k_{00}$, $k_{01}$, $k_{11}$) evaluated for KAg} along the trajectory shown in Fig.~\ref{fig3}(c). The right vertical axis in (a) displays the interaction strengths scaled by the interparticle distance $d \approx 4\,000\,a_0${, corresponding to a bulk gas density of $n = 10^{14}\,\mathrm{cm}^{-3}$}. The overall field strength is fixed at $\Omega^{\mathrm{tot}} = 10\times 2\pi\,\mathrm{MHz}$.}
    \label{fig:KAg}
\end{figure}

To investigate this, we performed coupled-channel scattering calculations for KAg as a representative example ($d= 8.5\,\mathrm{D}$ and $\alpha_{e} = 8.1\,\mathrm{MHz}$, compared to $d=4.6\,\mathrm{D}$ and $\alpha_{e}=6.9\,\mathrm{MHz}$ for NaCs). Evaluating the tunability of the model parameters and the two-body loss rate coefficients along the previously discussed $\Omega_{\sigma}/\Omega_{\pi} = 2$ trajectory (see Fig.~\ref{fig:KAg}), we find that the loss rates are again suppressed below the $10^{-13}\,\mathrm{cm}^3/\mathrm{s}$ level. Simultaneously, the strengths of the interaction parameters $V$, $J_z$, and $W$ are substantially enhanced---by factors ranging from two to six---reaching values on the order of $10^4\,a_0$. Such a length scale exceeds both the interparticle spacing of a bulk droplet and the lattice constant for both standard 1064~nm and visible-wavelength (e.g., 532 nm) optical lattices considered for such molecules \cite{Porsev2025, Vayninger2025}.
The enhancement is consistent with the expected $a_{dd} \propto Md^{2}$ scaling \cite{Dutta_2025}, which predicts a factor of $\approx 3.2$ based on the bare dipole moments and molecular masses. This demonstrates that the state-dependent shielding scheme can be extended even further into the strongly interacting regime without sacrificing stability.

\bibliography{bibliography}
\end{document}